%% file: main.tex
\documentclass[manuscript,authorversion,nonacm]{acmart}

\AtBeginDocument{%
  }

\setcopyright{acmlicensed}
\copyrightyear{2026}
\acmYear{2018}
\acmDOI{XXXXXXX.XXXXXXX}
\acmConference[TOSEM '26]{Make sure to enter the correct
  conference title from your rights confirmation email}{June 03--05,
  2018}{Woodstock, NY}
\acmISBN{978-1-4503-XXXX-X/2018/06}

\usepackage{hyperref}
\usepackage[switch]{lineno}
\usepackage{paralist}
\usepackage{url}
\usepackage{booktabs}
\usepackage{float}
\usepackage{cleveref}
\usepackage{amsthm}
\usepackage{listings}
\usepackage{threeparttable}
\usepackage{makecell}
\usepackage{xcolor}
\usepackage{caption}
\newcommand{\TODO}[1]{\textcolor{red}{#1}\GenericWarning{}{LaTeX Warning: TODO: #1}}\newcommand\todo\TODO
\usepackage{enumitem}

\crefname{lstlisting}{listing}{listings}
\Crefname{lstlisting}{Listing}{Listings}

\lstdefinelanguage{Nix}{
  keywords={rec, let, in, with, if, then, else, import, fetchFromGitHub, buildPythonPackage},
  sensitive=true,
  comment=[l]{\#},
  string=[b]",
  morestring=[b]'',
  alsoletter={-},
}

\definecolor{nixkeyword}{RGB}{133,153,0}
\definecolor{nixstring}{RGB}{42,161,152}
\definecolor{nixcomment}{RGB}{147,161,161}
\definecolor{nixattribute}{RGB}{38,139,210}
\definecolor{nixbackground}{RGB}{253,246,227}

\usepackage{minted}
\usepackage{xcolor}

\definecolor{codebg}{RGB}{253,246,227}

\setminted{
    fontsize=\footnotesize,
    breaklines=true,               %
    breakanywhere=true,            %
    frame=single,
    framesep=2mm,
    bgcolor=codebg,                %
    linenos=true,                  %
    numbersep=5pt,
    style=solarized-light          %
}
\usepackage{subcaption}
\begin{document}

\title{Vibenix: Agentic Software Packaging and Maintenance with Nix}

\author{Martim Monis}
\email{martimrosam@tecnico.ulisboa.pt}
\orcid{0009-0006-2175-1403}
\affiliation{
    \institution{INESC-ID and IST, University of Lisbon}
    \city{Lisbon}
    \country{Portugal}
}

\author{Martin Schwaighofer}
\email{martin.schwaighofer@ins.jku.at}
\orcid{0009-0001-1572-0495}
\affiliation{%
  \institution{Johannes Kepler University Linz}
  \city{Linz}
  \country{Austria}
}

\author{Nuno Saavedra}
\email{nuno.saavedra@tecnico.ulisboa.pt}
\orcid{0000-0003-4148-5991}
\affiliation{%
\institution{INESC-ID and IST, University of Lisbon}
\city{Lisbon}
\country{Portugal}
}

\author{Alexandra Mendes}
\email{alexandra@archimendes.com}
\orcid{0000-0001-8060-5920}
\affiliation{%
  \institution{INESC TEC, Faculdade de Engenharia, Universidade do Porto}
  \city{Porto}
  \country{Portugal}
}

\author{Rene Mayrhofer}
\email{rm@ins.jku.at}
\orcid{0000-0003-1566-4646}
\affiliation{%
  \institution{Johannes Kepler University Linz}
  \city{Linz}
  \country{Austria}
}

\author{João F. Ferreira}
\email{joao@joaoff.com}
\orcid{0000-0002-6612-9013}
\affiliation{%
  \institution{INESC-ID and Faculdade de Engenharia, Universidade do Porto}
  \city{Porto}
  \country{Portugal}
}

\renewcommand{\shortauthors}{Monis et al.}

\input{chapters/00_abstract}



\maketitle

\input{chapters/01_introduction}

\input{chapters/02_background}

\input{chapters/03_related_work}

\input{chapters/04_implementation}
\input{chapters/05_evaluation}

\input{chapters/06_conclusion}
\input{chapters/07_acknowledgments}

\input{chapters/A_additional_validation}

\bibliographystyle{ACM-Reference-Format}
\bibliography{references}

\end{document}

%% file: chapters/00_abstract.tex
\begin{abstract}

Implicit dependencies, uncontrolled network access, and other build-environment impurities create intransparent software supply chains. Identifying every dependency affecting a build is a difficult task. Functional package management addresses this through reproducible packaging, ensuring software relies exclusively on declared, hermetically isolated inputs. However, creating build recipes remains complex and time-consuming, posing a significant barrier to adoption for developers and maintainers.

We present Vibenix, an open-source AI-powered agent that bridges this gap by automatically generating Nix packaging expressions. Vibenix uses a large language model (LLM) to iteratively refine recipes through a rule-guided build-repair feedback loop. On 466 packaging tasks from our new NixBench dataset, Vibenix produced buildable packages in 95.3\% of cases using Claude Haiku 4.5, compared with 5.6\% for a state-of-the-art LLM-based single-shot packaging tool. 
Manual validation of 100 packages revealed that 47\% of the buildable packages were initially functionally correct;
a post-build refinement process based on the evaluator-optimizer pattern increased functional correctness by 35 percentage points. On our new NixBench-Repair dataset of package maintenance tasks, Vibenix achieved an 87.6\% success rate.

These results demonstrate how unreliable LLMs can generate rigorously defined, complete, auditable dependency trees for real-world software, reducing maintainer burden while providing a trustworthy foundation for secure software distribution.
\end{abstract}

%% file: chapters/01_introduction.tex
\section{Introduction}
Modern software development is built upon a rapidly expanding software ecosystem. Rather than being developed in isolation, applications rely on complex graphs of third-party dependencies, sophisticated build toolchains, and system-level libraries, resulting in unprecedented levels of complexity.
While this ecosystem accelerates development, it introduces the pervasive ``it works on my machine'' problem, where discrepancies between development, testing, and production environments lead to deployment failures and complex debugging scenarios. To mitigate these discrepancies, the industry has widely adopted software and hardware isolation techniques, such as containers and virtual machines.

However, runtime isolation alone is insufficient to guarantee the integrity of the software itself. The opacity of modern build systems has fueled a rise in software supply chain attacks, where adversaries compromise software at the source or inject malicious code during the build process~\cite{peisert2021perspectives}. A stark illustration of such an attack is the recent Axios supply chain attack~\cite{axios2026backdoor}, where a compromised maintainer account was used to publish \textit{backdoored} versions of the Axios software that included an additional malicious package dependency, which fetched a cross-platform remote access trojan to the host machine, all achieved via seemingly routine package updates. In standard build environments, dependency managers often automatically resolve loose version ranges and allow arbitrary code execution, such as \texttt{postinstall} scripts, with full network access to fetch secondary payloads. This creates non-deterministic behaviors that are difficult to audit, meaning that even isolated containers may end up deploying compromised artifacts.

To combat these threats, tools like Nix~\cite{dolstra2004nix} and Bazel~\cite{bazel2024} have emerged as the gold standard for \emph{hermetic packaging}, which consists in the practice of isolating package builds so they depend solely on explicitly declared inputs, independent of the host system's state. This ensures that software artifacts are bit-for-bit reproducible, providing a robust foundation for supply chain security while simultaneously eliminating a vast class of configuration errors.
The design principles which support these properties have also been used to offer performance benefits, namely through incremental builds, build parallelization and caching~\cite{zheng2024does, ranjan2025cost}.

Despite these critical advantages, the adoption of hermetic build systems remains largely hindered by several obstacles, that often include a steep learning curve. With regards to Nix in particular, via the 2024 Nix Community Survey~\cite{nix_community_survey_2024}, among respondents who do not consider Nix part of their usual toolset, the most commonly cited barriers were ``still learning'', ``barrier to entry is too high'', ``confusing documentation'', and ``the Nix language is an obstacle''. For many developers, the manual effort required to migrate legacy projects to hermetic build systems and to maintain them on the long term, outweighs the perceived benefits~\cite{ranjan2025cost}.

To reduce the expertise and manual effort required for adopting hermetic build systems, we present Vibenix, an AI agent for automating the creation and maintenance of Nix packages. Rather than relying on one-shot generation, Vibenix combines iterative execution-guided repair with post-build evaluation and optimization to produce correct and maintainable package definitions. To support a rigorous evaluation of this approach, we also introduce two benchmark datasets for packaging and maintenance tasks. The main contributions of this paper are as follows:

\begin{enumerate}
    \item \textbf{Vibenix:} an AI agent that automatically produces hermetic Nix expressions for real-world projects, using two key techniques: a log-driven loop that incrementally fixes build failures, and a post-build evaluator–optimizer process that improves correctness and package quality. Additionally, \textit{Vibenix} can also perform maintenance on pre-existing packages, ensuring long-term project support. We provide Vibenix as open source and all the data and code associated with our experiments.
    \item \textbf{NixBench and NixBench-Repair:} the first datasets for systematically evaluating automated Nix packaging and maintenance, comprising 472 and 760 curated, recent tasks, respectively, from the Nixpkgs repository. We additionally provide a detailed and reproducible methodology for constructing both datasets, enabling future extensions while mitigating data-leakage risks as time progresses and the training cutoffs of large language models continue to advance.
    \item \textbf{Empirical Evaluation of Agentic Packaging and Repair:} We evaluate Vibenix across both diverse datasets of open-source software packaging and maintenance tasks, and show that it can autonomously satisfy 95.3\% of these packaging requests, and fulfill 87.6\% of packaging repairs.
\end{enumerate}

To guide this work, we pose the following research questions (RQs):

\def\RQOne{To what extent can general-purpose LLMs successfully package diverse software projects using an iterative build-repair feedback loop?}
\def\RQTwo{To what extent does an LLM-driven refinement step improve the runtime correctness of generated packages, and at what cost?}
\def\RQThree{To what extent can an LLM-driven iterative build-repair approach reliably maintain real-world packages?}
\begin{itemize}
    \item \textbf{RQ1:} \RQOne
    \item \textbf{RQ2:} \RQTwo
    \item \textbf{RQ3:} \RQThree
\end{itemize}

All the instructions on how to replicate the evaluation experiments performed, and their respective data, is made entirely publicly available at \url{https://github.com/infragov-project/vibepkgs/tree/v2.0.0}.

Vibenix is open-source and publicly available at \url{https://github.com/infragov-project/vibenix}.

This paper significantly extends our preliminary work published at the 2026 Workshop on LLM Assisted Security and Trust Exploration (LAST-X)~\cite{schwaighofer2026vibenix} by introducing a new package-maintenance mode and associated dataset, improving the refinement phase, introducing a new automated validation metric, and substantially expanding and strengthening the empirical evaluation.

%% file: chapters/02_background.tex
\section{Software Packaging with Nix}
\label{sec:nix_example}

In this section, we briefly introduce the key concepts to understand the motivations and challenges of developing an AI agent for packaging software with Nix. First we analyze the process of software packaging with Nix, and then we cover relevant developments in Large Language Models (LLMs).

Hermetic packaging in Nix involves writing a functional Nix expression that declaratively specifies every build step of the complete dependency graph, which is then executed in the Nix build sandbox. This purely functional model ensures that every component is immutable and stored in a global location, the Nix store (\texttt{/nix/store}), at a path uniquely derived from a 160-bit hash of its entire input dependency graph.

\begin{listing}[htbp]
\centering
\begin{minted}{Nix}
{ lib
, python3Packages
, fetchFromGitHub
, gdal
}:

python3Packages.buildPythonPackage rec {
  pname = "stackstac";
  version = "0.5.1";

  src = fetchFromGitHub {
    owner = "gjoseph92";
    repo = "stackstac";
    rev = "v${version}";
    hash = 
     "sha256-7mll2jiZEx0xPDATWCBgqle4I68pkL6BeTLHKRmXEss=";
  };

  pyproject = true;

  build-system = with python3Packages; [
    pdm-pep517
  ];

  dependencies = with python3Packages; [
    xarray
    dask
    rasterio
    pystac-client
    numpy
    pyproj # Added missing dependency
    pandas # Added based on evaluator feedback
  ];

  buildInputs = [
    gdal
  ];

  doInstallCheck = true;
  installCheckPhase = ''
    ${python3Packages.python.interpreter} -c "import stackstac"
  '';
}
\end{minted}
\caption{Nix package definition for stackstac.}
\label{lst:stackstac}
\end{listing}

Listing~\ref{lst:stackstac} shows a functional Nix expression generated by Vibenix that packages a Python library called \texttt{stackstac}\footnote{\url{https://github.com/gjoseph92/stackstac}}.
In addition to the Nix language, the listed code heavily relies on the infrastructure code and dependencies provided by Nixpkgs\footnote{\url{https://github.com/nixos/nixpkgs}} to abstract away the details. This domain-specific, lazily evaluated language allows for defining complex, interdependent systems where values are computed only when required.
When using Nix like this, a package definition is structured as a function that accepts a set of arguments representing packages on which the definition depends and language-specific infrastructure code that supports the build (lines 1 to 5).

Line 7 introduces a function invocation with  a block that defines a Python package. Within this block, the package is assigned the name \texttt{stackstac} (line 8) and the version \texttt{0.5.1} (line 9).
In Nixpkgs, the \texttt{src} attribute of a package specifies the location of its source code. It typically references a file path, or a fetcher function that retrieves the source code of the package from a Git repository, or tarball URL. In our example, a fetcher function is then used to obtain the Python package's source code (line 11 to 17).

Lines 19 to 23 ensure that the build invokes the correct build infrastructure for Python projects provided in the \texttt{nixpkgs} repository.
From lines 25 to 37, project dependencies are explicitly declared. Among them, \texttt{gdal} represents a seamlessly integrated C/C++ dependency that originates outside the Python ecosystem. The accompanying comments were automatically inserted by Vibenix during various stages of the fully automated packaging process.
Finally, lines 39 to 42 introduce a build-time check that verifies that the package can be successfully imported without raising errors.
These dependencies are referenced solely by their name and are resolved from Nixpkgs. Because Nixpkgs pins package versions to specific revision commits to ensure reproducibility, all underlying dependencies remain fixed for that release. If a project requires a dependency version distinct from the one provided by the Nixpkgs release, it is necessary to override it using a Nix fetcher and explicitly declare its cryptographic hash.

Packaging software with Nix is generally more challenging than with other tools, primarily because many build systems implicitly rely on environment dependencies or do not manage the full transitive closure of their dependency graphs. Even language-specific package managers that offer some degree of hermetic isolation, such as \texttt{cargo} for Rust, typically include escape hatches that allow unresolved system-level dependencies.
In case of Rust for example, whatever gets executed inside a \texttt{build.rs} file is not covered by \texttt{Cargo.lock} and cannot be reasoned about within the corresponding dependency graph. As a result, such packages can easily evade software supply chain analyses, highlighting a fundamental limitation that our work seeks to overcome by easing the Nix packaging process.

Nix also enforces strict build-time purity by disallowing network access during builds unless all resources are accompanied by precomputed cryptographic hash values. This guarantees that all dependencies are explicitly declared, allowing users to fully understand, track and archive the set of dependencies of a package with confidence in its completeness.
From a security perspective, Adkins et al.~\cite{adkins2020building_chapter14} found that hermetically isolated builds facilitate the analysis and enforcement of policies on build inputs, ensure the integrity of third-party imports, and simplify the rapid deployment of critical software updates.

However, Nix also introduces unique challenges.
We already mentioned disallowing network access in the build sandbox, which requires extra effort and consideration during the packaging process.
The sandbox environment is also practically empty except for declared dependencies.
Additionally, Nix does not adhere to the Filesystem Hierarchy Standard (FHS), which often makes it necessary to patch source code that assumes dependencies reside in usual FHS-prescribed locations.
For example, a program may attempt to invoke an external binary using a hard-coded path such as \texttt{/usr/bin/git}, whereas under Nix the corresponding executable resides in the Nix store. Such assumptions can therefore lead to runtime failures in otherwise successful build outputs and must be identified and addressed iteratively during the packaging process.

Addressing these sandbox failures is a highly iterative, frustrating process. A developer must run the build, analyze the potentially obscure and lengthy log output, identify where the failure originates, apply the necessary patch, and rebuild. 
This tedious reasoning loop highlights the difficulty in adopting Nix at scale. However, it is also precisely what makes Nix packaging an ideal candidate for automation via LLMs and autonomous agents. By interpreting build logs, reasoning about the Nixpkgs infrastructure, and systematically applying fixes, agentic AI can navigate the strict rules of the Nix sandbox to automatically synthesize valid, reproducible package compositions.

%% file: chapters/03_related_work.tex
\section{Related Work}
\label{sec:related_work}

This section contextualizes \textit{Vibenix} by reviewing current packaging automation solutions, the challenges of low-resource languages, and the shift toward iterative, agentic repair workflows.

\subsection{Software Supply Chain Security and Hermetic Builds}

Hermetically isolated builds are widely recognized as a stepping stone towards reproducible builds~\cite{zheng2024build} and are supported by tools such as Bazel~\cite{bazel2024}, Guix~\cite{courtes2013functional}, Nix~\cite{dolstra2004nix} and Gitian\footnote{https://github.com/devrandom/gitian-builder}. Both concepts strengthen software supply chain security~\cite{enck2022top, lamb2021reproducible, malka2024reproducibility, fourne2023s}. For example, security researchers have explored hermetic builds to mitigate software supply chain security issues~\cite{lins2024critical,schwaighofer2023extending} and to maintain an indexed archive of binary software packages~\cite{schwaighofer2023extending}.

Beyond security, cloud-based build systems~\cite{BuildSystemsALaCarte}, which in addition to hermetically isolated builds provide output lookups by hash, like Nix, Guix and Bazel, have demonstrated their ability to enable reproducible builds at scale~\cite{zheng2024does, malka2025does, malka2024reproducibility}, and facilitate the creation of indexed archives of source code and binary packages to facilitate the reproducibility of science~\cite{10.1145/3641525.3663622} and aid practical debugging. However, despite these advantages, recent findings by Alfadel and McIntosh~\cite{alfadel2024classics} reveal a trend of organizations migrating away from Bazel due to unexpectedly high maintenance costs. Specifically, these costs manifest as a steep learning curve for developers, a persistent overhead of manually configuring granular dependency graphs, and the friction of navigating the restrictions inherent to a rigid, hermetic ecosystem. This positions our work as a means to address this challenge by lowering the maintenance burden associated with hermetic build systems.

\subsection{Automated Nix Packaging}

Traditional automation in Nix relies on ``2nix'' tools (e.g., \texttt{pip2nix}, \texttt{nix-init}, \texttt{uv2nix}) which use deterministic heuristics to generate expressions or derivations for specific languages. While each one is effective for projects built on an individual language, the majority of these tools lack the flexibility to handle cross-language dependencies, and require continuous maintenance from the developers to keep them updated on the specific ecosystem they handle. Furthermore, they are one-shot generators that do not introduce any form of output validation to their execution. \textit{Vibenix} differs by using an iterative LLM-based loop that reacts to real-time build failures.

\texttt{nix-ai-help}\footnote{\url{https://github.com/olafkfreund/nix-ai-help}} is an LLM-powered CLI tool providing various functionalities for Nix users, including automated project packaging through its \texttt{package-repo} command and build failure analysis through its \texttt{build} command. In comparison to Vibenix, \texttt{nix-ai-help} employs a single-shot LLM inference for package generation, lacking iterative improvements based on build output, a progress evaluation heuristic, model tools and an automated runtime validation. Moreover, \texttt{nix-ai-help} does not integrate build failure analysis with the project packaging process.

\subsection{LLM-based Code Generation}

LLMs have fundamentally transformed automated code generation by learning the semantic structure of programming languages from large-scale datasets of software repositories. Early models, such as Codex, demonstrated that transformer-based language models could synthesize executable code directly from natural language descriptions by using zero-shot program synthesis and multi-turn code generation, achieving promising performance on programming tasks written in general-purpose programming languages~\cite{chen2021evaluating,austin2021program,hendrycks2021measuring,nijkamp2022codegen}.

Rapid advances in both model architectures and training methodologies have substantially improved code generation performance. Benchmarks such as HumanEval~\cite{chen2021evaluating}, MBPP~\cite{austin2021program}, BigCodeBench~\cite{zhuo2024bigcodebench}, and Mercury~\cite{du2024mercury} consistently demonstrate that modern LLMs achieve high functional correctness on isolated programming tasks, aided by larger context windows, improved reasoning capabilities, and stronger adherence to programming language syntax. Recent surveys further show that open-weight code models have narrowed much of the performance gap with proprietary systems. For example, DeepSeek-Coder-V2 and Qwen2.5-Coder achieve HumanEval and MBPP results approaching those of Claude and GPT-family models while requiring significantly fewer parameters~\cite{jiang2026survey}.

However, benchmark performance should be interpreted cautiously. Many widely adopted evaluation datasets have remained unchanged for several years, increasing the likelihood that recent models have encountered portions of these benchmarks during pretraining~\cite{ramos2025large}. Consequently, very high pass@1 scores on HumanEval or MBPP increasingly measure memorization and general programming competence rather than the ability to solve novel software engineering problems. More importantly, these benchmarks primarily evaluate isolated functions instead of the repository-scale reasoning, tool usage, and iterative debugging required in professional software development.

This distinction becomes evident in software engineering benchmarks such as SWE-bench \cite{jimenez2023swe}. Early evaluations reported that state-of-the-art models solved fewer than 4\% of real GitHub issues, despite their strong performance on conventional code-generation benchmarks. Although modern agentic systems paired with frontier models now exceed 70\% success on SWE-bench Verified, performance remains substantially lower on the more demanding SWE-bench Pro benchmark, where success rates remain around 50\% even for current state-of-the-art systems \cite{introducingSWEBenchVerified,deng2025swe,sweBenchProLeaderboard}. These improvements are attributable not only to stronger foundation models but also to execution-time enhancements such as tool use, iterative repair, and autonomous reasoning loops.

Specialized programming domains are even more challenging. IaC frameworks and many configuration languages are implemented as domain-specific languages (DSLs)~\cite{fowler2010domain}, whose limited adoption results in substantially less training data than general-purpose languages \cite{joel2024survey}. These languages belong to the broader category of low-resource programming languages (LRPLs), where scarce training corpora and highly specialized semantics limit the transferability of knowledge acquired from conventional software repositories.

Joel et al.~\cite{joel2024survey} surveyed over one hundred studies on LLM-based generation of DSLs and LRPLs, concluding that existing approaches rely primarily on fine-tuning, prompt engineering, and iterative repair while suffering from the absence of standardized evaluation benchmarks. The authors further note that generic code-generation metrics such as BLEU and pass@k often fail to capture domain-specific correctness, motivating specialized validation pipelines.

IaC generation illustrates these challenges clearly. Kon et al. introduced IaC-Eval, the first benchmark specifically designed for Terraform configuration generation, and observed that GPT-4 achieved only 19.36\% pass@1 accuracy despite exceeding 80\% accuracy on conventional Python benchmarks~\cite{kon2024iac}. Their evaluation further demonstrated that semantic validation based on infrastructure resource graphs eliminated approximately half of the syntactically valid yet functionally incorrect outputs. These results indicate that semantic validation is critical for preserving the intended semantics of infrastructure configurations, as LLMs frequently generate code that is syntactically correct but fails to capture the domain-specific intent and constraints required for correct execution. Motivated by this limitation, Vibenix incorporates semantic validation into its refinement stage. Delgado et al.~\cite{delgado2026framework} also evaluated LLMs on several formal DSLs, including Alloy and OCL, reporting that even strong general-purpose models experienced significant drops in both syntactic correctness and semantic validity compared with Python, while contemporary open-weight models often produced unusable outputs.

Collectively, these studies demonstrate that success in mainstream code generation does not directly transfer to specialized software engineering domains. Instead, reliable generation for DSLs and IaC increasingly depends on domain-specific guidance, execution feedback, specialized validation, and iterative repair mechanisms. These observations directly motivate the design of Vibenix, which combines tool augmentation, execution-guided repair, and domain-specific reasoning to address the unique challenges of generating Nix expressions for hermetic software packaging.

\subsection{LLM-based Environment Setup}

The process of automated Nix packaging is specifically a problem of environment configuration. Recent research has begun to explore how agentic workflows can navigate the diverse and undocumented requirements necessary to create functional build environments.

One notable approach to automating this operational layer is \textit{ExecutionAgent}~\cite{bouzenia2025you}. It introduces a general-purpose LLM-based agent, following a tool-augmented agent approach, that is capable of setting up environments and executing test suites across diverse languages and build systems. Evaluated on a total of 50 different open-source projects, it was able to successfully execute the test suite of 33 projects (a 66\% success rate), while generally matching the output from the ground truth samples, and taking around 70 mins and a lowly average of 0.16 USD per project.

As another relevant work, Eliseeva et al.~\cite{eliseeva2025envbench} propose EnvBench, a benchmark focused specifically on environment setup, covering 329 Python and 665 JVM-based repositories. They evaluate several LLM-based strategies and find that even the best agent successfully configures only 6.69\% of Python and 29.47\% of JVM repositories, highlighting the difficulty of the task and the limitations of current techniques.

Distinct from existing solutions, Vibenix advances supply chain security by automatically enforcing hermetic builds, closing off attack vectors that arise from non-deterministic and environment-dependent builds.

%% file: chapters/04_implementation.tex
\section{Vibenix}
\label{sec:implementation}

\begin{figure*}[t]
    \centering
    \includegraphics[width=0.80\textwidth]{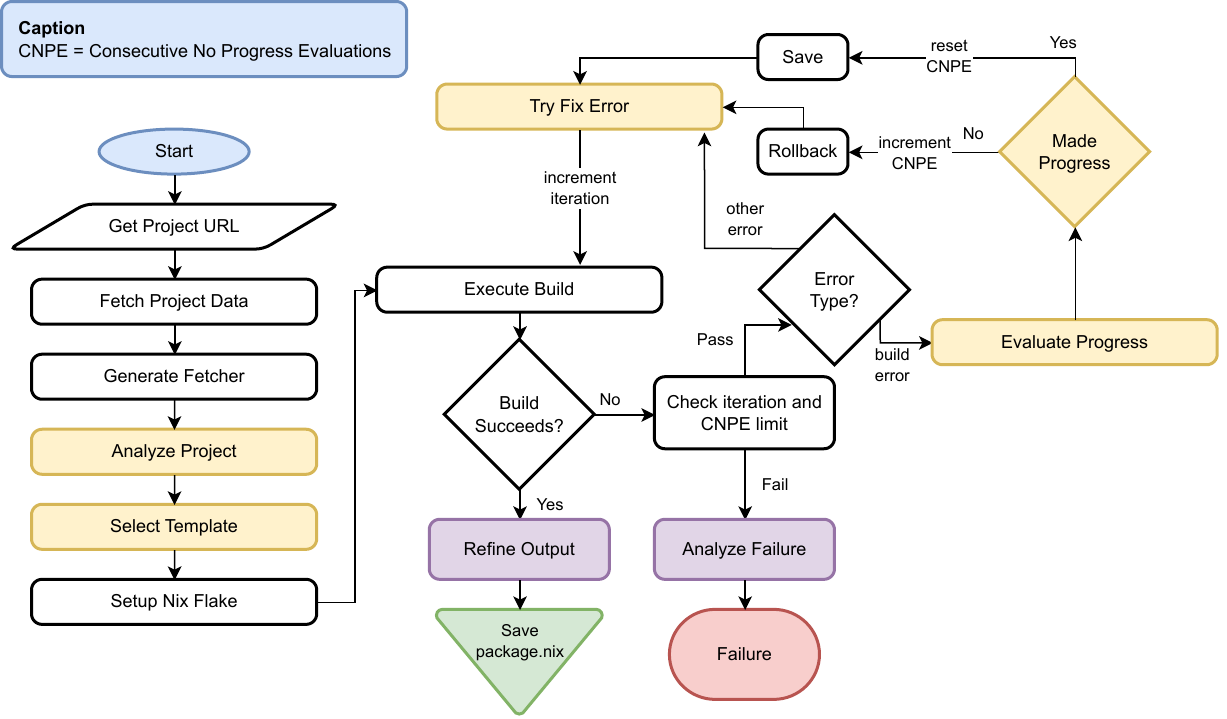}
    \caption{Overview of Vibenix's packaging pipeline. Yellow and purple boxes indicate use of an LLM in these steps.}
    \label{fig:packaging-flow-chart}
\end{figure*}
Vibenix employs an agentic architecture that leverages an LLM to support the fully automated packaging process.
As already explained in Section~\ref{sec:nix_example}, the problem is a natural fit for the application of such an LLM-based design.

Particularly, Nix is a good tool for applying LLMs to the problem of software packaging because it provides a mechanism to clearly scope tasks while enabling iteration within the narrowly defined scope of a package build over the entire build process at once.

Both of these properties make auditing and iteration on tasks easier than, for example, general-purpose coding tasks or imperatively setting up a build environment step by step.
The roughly linear nature of build processes also makes it theoretically easier than for general-purpose coding tasks to distinguish working solutions from failing solutions and judge whether the packaging process is moving in a productive direction.
The final output of the process is also more straightforward to validate than for a general-purpose coding task, since making an existing application work is a more well-defined goal.

\subsection{Packaging Pipeline}

In this section, we provide a step-by-step description of Vibenix's core execution flow. While the system supports two distinct execution modes consisting of initial package generation and package maintenance, this section focuses strictly on the former. However, much of the underlying process is identical between the two distinct execution modes, as explained later in Section~\ref{sec:maintenance_mode}. Figure~\ref{fig:packaging-flow-chart} presents an overview of this sequential process.
The process begins when the user provides the URL of the target project to Vibenix, the only required input from the user to generate a new package.

\subsubsection{Generate Fetcher}
\label{sec:fetcher}

To fetch the source code of the project, Vibenix uses a Nix fetcher invocation. These fetcher invocations are used to download and unpack the project source code from the specified input location, usually a remote git forge such as GitHub or GitLab, to the local nix store. To generate this fetcher invocation, our tool relies on \texttt{nurl}\footnote{\url{https://github.com/nix-community/nurl}}, which fetches supported sources with Nix and turns them into a call to a suitable fetcher function with the respective content hash already filled in. The expression assigned to the \texttt{src} attribute in Listing~\ref{lst:stackstac} is an example of a fetcher invocation generated by \texttt{nurl}.

Vibenix uses the local Nix installation to execute the generated fetcher, which clones the project source code into the local Nix store. The resulting store path is recorded for use within Vibenix, allowing it to retrieve relevant information about the project during subsequent stages of the packaging pipeline.

\subsubsection{Analyze Project}
\label{sec:summary}

Vibenix then prompts the LLM with information gathered from the project's source code, namely the project's root \texttt{README} file and root directory listing, to generate a summary that includes build tools, dependencies, and other details relevant to the build process. This prompt provides the model with tools to further analyze the project source at will.

\subsubsection{Select Template}

Rather than providing the LLM with a blank canvas, Vibenix guides its output by initializing it with a build tool or language-specific template that adheres to the style conventions of the \texttt{nixpkgs} repository.
To select the appropriate template, Vibenix prompts the LLM using the previously generated summary to identify the most suitable starting point from a list of several template names, identifying the underlying framework or language of the project to be packaged with. If no specific template can be matched to the target project, then a generic one is used.

After selecting the appropriate template, Vibenix populates its \texttt{src} attribute with the previously generated fetcher invocation.

\subsubsection{Setup Nix Flake}

Then Vibenix sets up a temporary project directory based on the selected template with both an additional \texttt{flake.nix} and \texttt{flake.lock} file, which provide a defined structure in terms of how the project build is invoked and which version of the existing \texttt{nixpkgs} package set it is based on. Vibenix takes advantage of existing package definitions from the \texttt{nixpkgs} repository, thereby eliminating the need to independently repackage each component required to build and run the project.

\subsubsection{Execute Build}

After the Nix flake is set up, Vibenix enters an agentic loop. This agentic loop makes multiple calls to the LLM until a successful build is reached or Vibenix reaches one of its stopping conditions.

First, Vibenix uses the local Nix installation to build the current Nix expression and collect the build results. To maintain a structured history of the packaging process, every attempt is stored in a \textit{solution stack}, which provides the LLM with the context necessary for reverting to previous state when it gets to a regression.

\subsubsection{Stopping Conditions} 
\label{subsec:stop_conditions}

If the build succeeds, the package optionally goes through a refinement process (see Section~\ref{subsec:refinement}), and then the final package is saved, and execution ends.

If the build fails, Vibenix evaluates whether one of its stopping conditions have been met, namely if \textit{the maximum number of allowed iterations has been exceeded}.
If one of the conditions is met, Vibenix stops its agentic cycle and outputs a failure report (see Section \ref{sec:failure_analysis}). Otherwise, the iterative process continues.

\subsubsection{Handling Errors}

If the agentic cycle continues, Vibenix inspects the error that caused the failure and attempts to fix it. Depending on the stage at which the failure occurred, the error is classified as either a \emph{build error} or an \emph{evaluation error}. A build error occurs when the LLM generates a valid Nix expression that successfully evaluates into a dependency graph but fails during the build phase. Otherwise, the failure is classified as an evaluation error. If the \textit{Evaluate Progress} step is enabled, Vibenix first assesses whether progress has been made (see Section~\ref{sec:progress_evaluation}) before attempting to fix build errors.

\subsubsection{Evaluate Progress}
\label{sec:progress_evaluation}

For each build error, Vibenix optionally evaluates whether a build failure represents progress over a previous attempt. When disabled, it is identical to considering all build errors as progress.

When enabled, the system compares the current build log with the last recorded failure in the solution stack. To optimize the model's context window, logs are preprocessed: if they exceed 100 lines, Vibenix identifies the first divergent line between the two logs and discards all preceding content. This heuristic approximates the divergence point in non-sequential parallel build logs. The LLM then analyzes the truncated logs to output a status: \texttt{PROGRESS}, \texttt{STAGNATION}, or \texttt{REGRESS}.

If \texttt{PROGRESS} is detected, the attempt is saved and the \textit{number of consecutive evaluations without progress} counter resets. Otherwise, Vibenix increments the counter and performs a \textbf{rollback} to the last known-good state. This mechanism prevents the agent from pursuing unproductive repair branches and exhausting computational resources when it fails to advance within the packaging process.

\subsubsection{Try Fix Error}
\label{sec:try_fix_error}

In this step, Vibenix prompts the LLM to fix the error in the current Nix expression.
Vibenix provides the LLM with an input consisting of the current Nix expression, the error logs, and the summary generated in Section~\ref{sec:summary}.

Vibenix also provides the model several tools, namely:
\begin{enumerate}
    \item \textbf{Edit tools}:
    Directly returning the entire packaging expression, or even snippets, at the end of error fixing prompts can often introduce noise, especially surrounding long strings like source hashes, and increased costs, considering output tokens are generally quite expensive (upwards of 5 times more than input tokens). For this reason, the model can have access to standardized tools to perform localized edits to the packaging expression and view its current state.
    \item \textbf{Package Search}: Given a query, the tool retrieves relevant packages from the \texttt{nixpkgs} repository by performing a search on all the packages available.
    \item \textbf{Find Nix Package Output}: Given a relative file path, the tool retrieves packages in Nixpkgs that provide this file path in their output. Leverages the \texttt{nix-locate} command.
    \item \textbf{File-centric tools}: 
    Tools that interact with the file system, including listing directory contents, reading file contents, and performing pattern-based searches. These operations are supported in both the project source directory and a commit of the \texttt{nixpkgs} repository.
\end{enumerate}

In the case of hash mismatch errors, Vibenix limits the input of the LLM to the current Nix expression and the error logs. For such scenarios, it employs a specialized prompt that includes guidance tailored to resolve this specific class of issues. Hash mismatches occur when using an unknown content hash of a network resource, such as a GitHub repository. Since Vibenix does not know correct hashes in advance, it fills in a dummy hash value (\texttt{lib.fakeHash}).
Subsequently, Nix will return a hash mismatch error, which includes the correct hash value, which Vibenix will then fix.
This trust-on-first-use workflow mirrors best practice in manual Nix packaging~\cite{hashNixpkgsmanual}.

\subsubsection{Refinement}\label{subsec:refinement}

\begin{figure*}[t]
    \centering
    \includegraphics[width=0.85\textwidth]{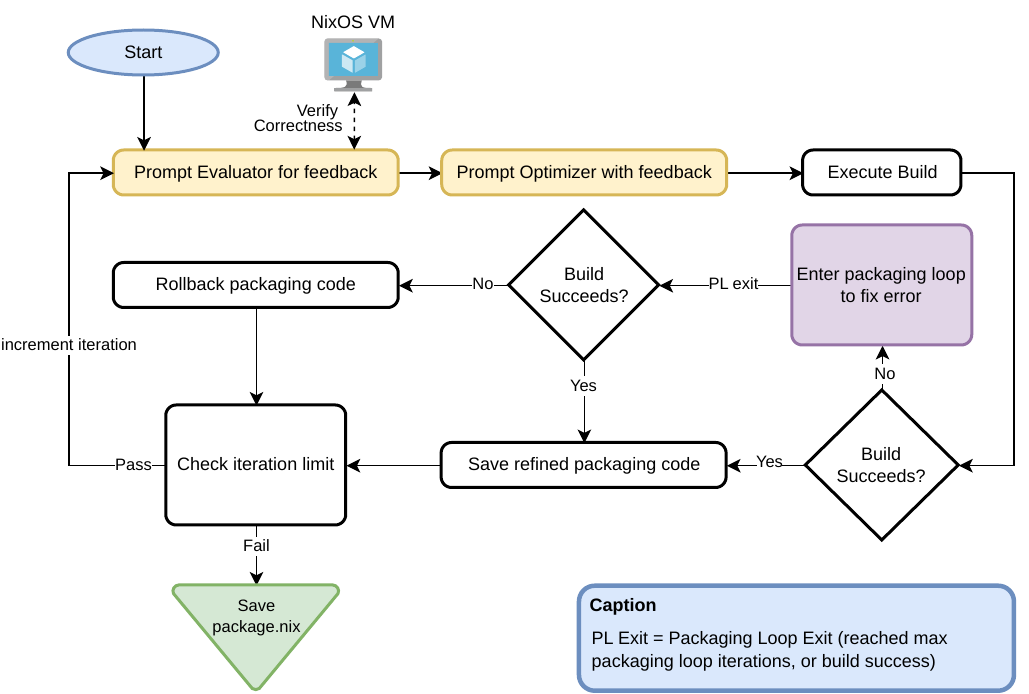}
    \caption{Overview of Vibenix's refinement stage. Yellow and purple boxes indicate use of an LLM in these steps.}
    \label{fig:refinement-flow-chart}
\end{figure*}

Once an initial build is successfully produced, either through the first build attempt or during the main agentic loop, which we named \textit{packaging loop}, the core goal is considered to be reached. However, Vibenix also includes a next, optional, \textit{Refinement} step. This step is specifically designed to improve package quality, presentation, and detect and correct errors not detectable during build. Its implementation draws inspiration from the agentic workflow pattern of the Evaluator-optimizer~\cite{anthropic2024}. Figure~\ref{fig:refinement-flow-chart} presents an overview of this process.

Each iteration of the refinement process primarily involves two prompts to the LLM. The first, the \textit{Evaluator} prompt, is provided with the Nix expression and relevant project information, and is tasked with identifying and describing possible improvements in packaging. Because the package received by the \textit{Evaluator} always builds, it can use the \texttt{run\_in\_vm} tool, which allows the LLM to specify and execute shell scripts to run in an isolated NixOS Virtual-Machine (VM). This tool allows Vibenix to autonomously verify functional correctness, inspect the outputs produced, and in this way increase the likelihood that the final packaging expression is sound.

After the \textit{Evaluator} replies with feedback, the \textit{Optimizer} is tasked with implementing the suggested changes. If the code produced by the \textit{Optimizer} successfully builds, it replaces the previous expression. Otherwise, it goes through the packaging loop for a pre-defined number of iterations, in an attempt to fix the errors introduced. From this inner loop, the package is either built successfully once more, or the maximum number of error fixing iterations is reached and the changes are discarded.
Regardless of success, the refinement process continues until a predefined maximum number of iterations is reached. Once it concludes, the final packaging code is one produced during this process or the one initially provided to it.

To optimize costs and context, Vibenix maintains a condensed chat history containing only successful code iterations and summarized error resolutions.

\subsubsection{Failure Analysis}\label{sec:failure_analysis}

If, by the end of the agentic loop, Vibenix fails to produce a successfully building Nix expression, it goes through a last failure analysis step.

In the first prompt, it receives the packaging code, the most recent build error that led to progress, and relevant project context. Then a concise explanation for the failure is asked. In the second prompt, the LLM is tasked with classifying the generated explanation into one of several predefined categories, such as \textit{dependency not in Nixpkgs}, \textit{missing lock file}, or \textit{build downloads from network}.

The failure analysis description and the failure type are presented to the user.
This feedback provides insight into the root cause of the packaging issue, enabling the user to better understand the problem and then manually intervene.

\subsection{Maintenance}\label{sec:maintenance_mode}

While the core functionality of \textit{Vibenix} focuses on the generation of Nix expressions from scratch, its utility is inherently limited by the inevitable deprecation of the generated packages.

While a Nix package remains functional indefinitely if left untouched, maintenance is essential to keep software secure, modern, and compatible with the broader ecosystem. This ongoing process is triggered by upstream version releases, the availability of newer dependencies, and the release of critical security patches. To integrate these updates, a maintainer must systematically align the environment used to build the package, often by updating a \texttt{.lock} file or migrating to a newer nixpkgs release, and sync the source attributes of the project and cryptographic integrity hashes. However, depending on the changes introduced to the project or the environement, maintenance requires an iterative build-repair process identical to initial packaging, where the maintainer continuously adjusts dependencies, compilation flags, and build phases to resolve breaking changes.

Consequently, to address the long-term viability of the generated packages, \textit{Vibenix} also includes a \textit{Maintenance} mode designed to update and repair existing Nix packages.

This alternative execution mechanism does not require a radically distinct pipeline. It begins by ingesting the existing \texttt{package.nix}, and optionally the flake files. If a version update is requested, \textit{Vibenix} leverages the \textit{nix-update}\footnote{\url{https://github.com/Mic92/nix-update}} tool to synchronize version strings and recalculate cryptographic hashes. 
The updated project is then funneled into the standard packaging loop. Because the starting state is a previously functional package, the LLM is prompted with a higher level of confidence in the existing code, focusing on localized fixes rather than radical restructuring. The agent can also have access to a specialized \texttt{upgrade} tool that allows the model to dynamically upgrade the locked \texttt{nixpkgs} release when it detects required dependencies are missing from the used one. This allows for a ``lazy upgrade'' path that ensures environmental stability while overcoming version mismatches.

With this execution mode, \textit{Vibenix} evolves from simply a once-use initial generation tool into a tool capable of long term package maintenance.

\subsection{Technical Implementation}
\label{sec:technical_implementation}

\textit{Vibenix} is implemented as a Python-based CLI using the popular \texttt{pydantic-ai}\footnote{\url{https://github.com/pydantic/pydantic-ai}} framework. It integrates directly with the local Nix daemon to execute sandboxed builds. All LLM prompts and Nix packaging recipes are stored as \texttt{Jinja2} templates to ensure modularity and reproducibility across different model providers, available in our project repository.

\paragraph{Configuration}
\label{subsec:configuration}

Vibenix has configurable settings to change its execution behavior.
The program loads these options through environment variables and configuration files stored in a local \texttt{.vibenix} directory, with the former taking precedence. There are three distinct configuration groups that Vibenix loads, namely a \texttt{config.json} file to set-up the model and model provider to use, a \texttt{model\_settings.json} to specify model-specific configuration (e.g., temperature), and a \texttt{vibenix\_settings.json} to configure Vibenix-specific behavior. 
Regarding the default Vibenix-specific application configuration, the relevant file is provided on the project repository under the \texttt{defaults/} directory\footnote{\url{https://github.com/infragov-project/vibenix/blob/main/src/vibenix/defaults/vibenix_settings.py}}, and it has all the features except \emph{Refinement}, \emph{Compare Template Builders}, \emph{Search Nixpkgs Package Semantic}, \emph{Search Nixpkgs Documentation}, and \emph{Find Similar Builder Patterns} enabled.
These features were disabled because preliminary experiments showed that they incurred substantial computational costs, particularly \emph{Refinement} and \emph{Compare Template Builders}, while providing only marginal improvements in build success rate.

%% file: chapters/05_evaluation.tex
\section{Evaluation}
\label{sec:evaluation}

In this section, we describe the datasets used to evaluate \textit{Vibenix}, followed by an explanation of the evaluation methodology.

\subsection{Datasets: NixBench and NixBench-Repair}
To evaluate \textit{Vibenix}, we constructed two datasets from the Nixpkgs package repository\footnote{\url{https://github.com/nixos/nixpkgs}}, which is the largest open-source software monorepo as tracked by Repology\footnote{\url{https://repology.org/}}, with 129,000 packages and 4,257 maintainers at the time of writing. Using commits authored between August 1 and November 20, 2025. The August 1st start date prevents data leakage, as models used for evaluation were trained only on prior data, while November 20th represents our data collection cutoff.

\paragraph{NixBench (Generation Tasks)} We identified 1,146 commits introducing new packages. To ensure a rigorous baseline, we filtered for commits exhibiting the basic structure expected of conventional Nixpkgs package additions: definitions must exceed six lines, include standard \texttt{meta}, \texttt{pname} and \texttt{version} attributes, and a single, valid Nix fetcher. We excluded commits that introduced multiple packages, that touched or introduced multiple files, packages requiring source-code patches, those that require special composition via \texttt{all-packages.nix}, and those marked as \textit{broken} or \textit{unfree}. We also filtered for commits with messages that followed standard Nixpkgs convention to obtain their Fully Qualified Name (FQN). Subsequently, to avoid data leakage, for each commit we identified the immediately preceding Nixpkgs commit from a list restricted to the selected time frame. Then, we evaluate the corresponding FQN using the selected Nixpkgs commit, and if successful, we conclude by performing a build with a timeout limit of 15 minutes. Only packages that successfully built on the respective Nixpkgs base were retained. The final NixBench dataset contains 472 valid generation tasks.

\paragraph{NixBench-Repair (Maintenance Tasks)} We sampled 21,274 commits modifying an existing \texttt{package.nix} or \texttt{default.nix} file within the same time frame as NixBench. To retain isolated package updates, we excluded commits modifying multiple packages, touching \texttt{all-packages.nix}, or modifying auxiliary files. We then applied the same baseline structural criteria as for NixBench, requiring package definitions to exceed six lines, include the standard \texttt{meta}, \texttt{pname}, and \texttt{version} attributes, and use a valid Nix fetcher. To focus on non-trivial maintenance rather than superficial updates, we further required a functional diff of at least three lines, excluding changes to \texttt{meta}, \texttt{src}, \texttt{version}, and comments. Finally, we obtained the FQN and selected an appropriate Nixpkgs base commit using the same reference window as for NixBench, before applying the same evaluation and buildability constraints. This process resulted in 760 maintenance tasks comprising the original \texttt{package.nix}, the Nixpkgs base commit, the target version, and the human-authored reference solution.

\subsubsection{Dataset Diversity}
To evaluate how effectively \textit{Vibenix} generalizes across different software ecosystems, we analyzed the distribution of programming languages and build frameworks within both final datasets. These statistics were obtained via simple textual regular expression matching over each package (e.g., detecting attributes such as \texttt{buildPythonApplication}, \texttt{buildGoModule}). Consequently, while these metrics provide a representative overview of the ecosystem layout, they are limited in absolute accuracy.

Table~\ref{tab:dataset-language-diversity} outlines the distribution of the most prominent ecosystems across the tasks in \emph{NixBench} ($N=472$) and \emph{NixBench-Repair} ($N=760$).

\begin{table}[H]
\centering
\caption{Distribution of programming languages and ecosystem frameworks across the final evaluation datasets.}
\label{tab:dataset-language-diversity}
\small
\begin{tabular}{l c c}
\toprule
\textbf{Ecosystem} & \textbf{NixBench} & \textbf{NixBench-Repair} \\
\midrule
Python & 253 (53.6\%) & 466 (61.3\%) \\
Rust & 47 (10.0\%) & 68 (8.9\%) \\
Go & 40 (8.5\%) & 73 (9.6\%) \\
Node & 7 (1.5\%) & 16 (2.1\%) \\
Mixed & 10 (2.1\%) & 22 (2.9\%) \\
Other* & 115 (24.3\%) & 115 (15.1\%) \\
\midrule
\textbf{Total} & \textbf{472 (100.0\%)} & \textbf{760 (100.0\%)} \\
\bottomrule
\end{tabular}
\end{table}

The structural composition demonstrates that both datasets are representative of modern open-source development trends, with Python, Rust, and Go accounting for the majority of tasks. Ecosystems with very few entries (fewer than five) were merged into the ``Other'' category, which is primarily composed of C/C++ projects. Due to the matching approach used, along with the fact that C/C++ projects heavily leverage the generic \texttt{mkDerivation} build function, further isolating and distinguishing these projects from the remaining miscellaneous group is an impractical task.

\subsection{Evaluation Methodology}
\label{sec:evaluation_methodology}

Experiments were conducted via GitHub Actions, with each packaging task run in isolation as a dedicated job. The models chosen for evaluation and their respective training cutoff dates are reported in Table~\ref{tab:model_cutoffs}. Importantly, all training cutoffs precede the dates of the commits in our datasets to prevent data leakage.

\paragraph{Sampling Methodology}
For all of our evaluation experiments except for full dataset evaluation, we use a statistically representative subset of 267 packaging tasks from the NixBench dataset. Following Cochran's sample size formula~\cite{ahmed2024choose}, this number ensures a maximum 6\% margin of error at a 95\% confidence level.
We excluded six packages involving unsupported binary formats (\texttt{.jar}, \texttt{.ttf}) identified during early experiments, all leading to immediate execution failure.

\paragraph{Evaluation Pipeline}
Our evaluation follows a structured, seven step pipeline:

\begin{enumerate}
    \item \textbf{Baseline Model Comparison.} We begin by comparing the various models with all of \textit{Vibenix}'s tools and optional mechanisms disabled. This establishes a performance baseline and identifies the top-performing models. 
    
    \item \textbf{Feature Ablation.} Using the most economical model of the top-performing models, we assess the contribution of \textit{Vibenix}'s individual components. We start with the initial configuration described in Section~\ref{subsec:configuration}. We disable each remaining feature in isolation to assess their impact.
    
    \item \textbf{Full Configuration Model Comparison.} We conduct a second round of model comparisons, this time with the ablated, tool-enabled \textit{Vibenix} configuration obtained from the previous step.
    
    \item \textbf{Full Dataset Evaluation.} Having identified the optimal configuration and the best models, we execute a full-scale evaluation across the entire generation dataset, using the top-performing closed- and open-source models.
    
    \item \textbf{Evaluating Runtime Correctness (Refinement).} After optimizing the configuration for the best build success rate, we enable the \emph{Refinement} step for 3 iterations to evaluate its impact on improving package quality and functional correctness. We perform manual validation of a subset of 100-packages. This validation is performed within an isolated VM, inside of which both the unrefined and refined versions of the generated package are available, alongside the reference package. We leverage the standard \texttt{diff} utility for text-based evaluation of files, alongside \texttt{diffoscope}. Where applicable, a limited manual runtime inspection is also performed.
    
    \item \textbf{Maintenance Phase.} We evaluate \textit{Vibenix}'s secondary execution mode for package maintenance on a similarly sized 267-package subset from NixBench-Repair, with \emph{Refinement} enabled for 2 iterations to cover runtime-specific updates. 

    \item \textbf{nix-ai-help Comparison.} We evaluate the \texttt{nix-ai-help} tool on \textit{NixBench} to compare Vibenix to the success rates and costs of a single zero-shot generation approach. We choose \texttt{nix-ai-help} because it is the state-of-the-art LLM-based packaging tool for Nix.
\end{enumerate}

All experiments were subject to a strict 7 minute individual build time limit, set via Vibenix, a soft 60 minute total execution time limit, also set via Vibenix, and a hard 65 minute time limit configured through GitHub Actions.

\paragraph{Automated Validation}

To verify the generated expressions, we perform an additional autonomous validation of the results obtained with a custom script. For each package, the script:

\begin{enumerate}
\item Obtains the list of package validation related attributes set in its corresponding reference package. These attributes include, for example, \texttt{pythonImportsCheck}, and \texttt{doInstallCheck}. If no validation is set, the update is not considered for automated validation. Otherwise, the script continues.
\item Creates an overlay for the generated package, to use all attribute values obtained in the previous step from its reference package. This ensures that all package validation configured by package maintainers is instead used to validate Vibenix's generated version. 
\item Builds the overlayed package in its appropriate Nixpkgs base commit, executing all primary validation checks.
\item Builds the overlaid package's \texttt{passthru.tests} attribute, thereby executing any additional tests specified by maintainers that are not normally run. In the tables presented in Section~\ref{sec:results}, these outcomes are reported under \emph{Additional}.
\end{enumerate}

This autonomous validation is limited in scope because it relies on the existence of validation checks added by real maintainers to the reference packaging code. Of the 267 packages in the packaging evaluation subset, 200 (75\%) included some form of validation in their reference, compared with 228 (85\%) in the equally sized maintenance evaluation subset.

\section{Results}
\label{sec:results}

Now we present the results for each evaluation step, identifying relevant changes and performance differences. 

\subsection{Baseline and Optimized Results}

To establish a baseline, we first evaluated the complete list of models with all \textit{Vibenix} configurable features disabled, except the feature that allows the model to inspect project source code, which we deemed necessary to ensure the results were relevant. Baseline success rates ranged from 52.8\% (GPT-5-nano) to 85.8\% (Claude Haiku 4.5 and GPT-OSS-120B), demonstrating that the core iterative feedback loop alone is largely capable of achieving successful builds. Table~\ref{tab:model_comparison_disabled} reports the full results.

\begin{table}[htpb]
\centering
\small
\caption{Baseline model performance comparison.} %
\label{tab:model_comparison_disabled}
\begin{tabular}{l c c r r c c}
\toprule
& & & \multicolumn{2}{c}{\textbf{Cost (\$)}} & \multicolumn{2}{c}{\textbf{Automated Validation}} \\
\cmidrule(lr){4-5} \cmidrule(lr){6-7}
\textbf{Model} & \textbf{Success Rate} & \textbf{Total Time} & \textbf{Avg} & \textbf{Total} & \textbf{Base} & \textbf{Additional} \\
\midrule
\textbf{GPT-5-mini} & 85.4\% $\pm$ 4.2\% & 28:05:19 & 0.03 & 8.22 & 77.0\% $\pm$ 5.1\% & 75.5\% $\pm$ 5.2\% \\

\textbf{GPT-5-nano} & 52.8\% $\pm$ 6.0\% & 55:34:34 & 0.03 & 7.63 & 37.0\% $\pm$ 5.8\% & 35.5\% $\pm$ 5.7\% \\

\textbf{GPT-OSS-120B} & 85.8\% $\pm$ 4.2\% & 17:30:22 & 0.02 & 4.66 & 79.0\% $\pm$ 4.9\% & 76.5\% $\pm$ 5.1\% \\

\textbf{Claude Haiku 4.5} & 85.8\% $\pm$ 4.2\% & 21:40:29 & 0.14 & 47.92 & 77.0\% $\pm$ 5.1\% & 75.5\% $\pm$ 5.2\% \\

\bottomrule
\end{tabular}%
\end{table}

Then, we performed feature ablation using GPT-OSS-120B, the most economical of the two best performers from the baseline evaluation. 
Table~\ref{tab:ablation_comparison} presents the results obtained. \textit{Nixpkgs Tools} and \textit{Template Selection} proved critical, as disabling them led to notable drops in success and increased costs.
Conversely, several features introduced unexpected significant overhead without accuracy gains. Although designed to prune unproductive search branches, \textit{Evaluate Progress} surprisingly acted as a major temporal bottleneck, as its removal reduced packaging time by around 24\% (over six hours). Although this increased API costs by around 13\%, the time savings justified its removal in our context. However, the feature remains relevant in scenarios that prioritize cost reduction over speed or have lower success rates, as it enables unsuccessful attempts to terminate earlier. Removing \textit{Get Builder Functions}, which provides the model with an updated list of build helper function names, also reduced the packaging time by 32\%. Most unexpectedly, the \textit{Edit Tools} significantly inflated both packaging time (28\%) and costs (47\%) without improving success rates, indicating that their reasoning complexity outweighs their utility for this task.
Consequently, our optimized configuration builds on our initial configuration, but disables these three specific features that negatively impacted our results.

\begin{table}[htpb]
\centering
\small
\caption{Key feature ablation results using the GPT-OSS-120B model.}
\label{tab:ablation_comparison}
\begin{tabular}{l c c r r c c}
\toprule
& & & \multicolumn{2}{c}{\textbf{Cost (\$)}} & \multicolumn{2}{c}{\textbf{Automated Validation}} \\
\cmidrule(lr){4-5} \cmidrule(lr){6-7}
\textbf{Configuration} & \textbf{Success Rate} & \textbf{Total Time} & \textbf{Avg} & \textbf{Total} & \textbf{Base} & \textbf{Additional} \\
\midrule
Initial Config & 89.1\% $\pm$ 3.7\% & 25:49:12 & 0.04 & 10.66 & 79.5\% $\pm$ 4.8\% & 77.0\% $\pm$ 5.1\% \\
No Nixpkgs Tools & 83.1\% $\pm$ 4.5\% & 22:01:37 & 0.05 & 12.06 & 75.5\% $\pm$ 5.2\% & 73.0\% $\pm$ 5.3\% \\
No Pick Template & 85.4\% $\pm$ 4.2\% & 32:53:05 & 0.05 & 14.20 & 76.5\% $\pm$ 5.1\% & 75.0\% $\pm$ 5.2\% \\
No Progress Eval & 88.8\% $\pm$ 3.8\% & 19:31:29 & 0.04 & 12.19 & 81.0\% $\pm$ 4.7\% & 78.5\% $\pm$ 4.9\% \\
No Get Builder Funcs & 90.3\% $\pm$ 3.6\% & 17:29:10 & 0.04 & 11.33 & 80.5\% $\pm$ 4.8\% & 78.0\% $\pm$ 5.0\% \\
No Edit Tools & 87.6\% $\pm$ 4.0\% & 18:33:38 & 0.02 & 5.63 & 83.5\% $\pm$ 4.5\% & 81.0\% $\pm$ 4.7\% \\
\bottomrule
\end{tabular}%
\end{table}

Using this optimized configuration, all models showed substantial gains over their baseline (Table~\ref{tab:model_comparison_final}). GPT-5-nano exhibited the greatest improvement (+27\% success), highlighting the possibility that lightweight, yet capable models most significantly benefit from tool-augmentation. Claude Haiku 4.5 reached 94.0\% build success via exhaustive tool usage, while the open-source GPT-OSS-120B performed remarkably close to the other proprietary models. 

When considering automated validation with the additional tests, success rates range from 63\% to 86\%. GPT-5-mini exhibits the largest performance decrease, with a drop of approximately 19.3 percentage points relative to its build success rate. Although GPT-5-mini achieves a build success rate comparable to Claude Haiku 4.5, a larger proportion of its successful builds fail the validation tests, indicating lower functional correctness.

\begin{table}[htpb]
\centering
\small
\caption{Final model comparison with optimized configuration.}
\label{tab:model_comparison_final}
\begin{tabular}{l c c r r c c}
\toprule
& & & \multicolumn{2}{c}{\textbf{Cost (\$)}} & \multicolumn{2}{c}{\textbf{Automated Validation}} \\
\cmidrule(lr){4-5} \cmidrule(lr){6-7}
\textbf{Model} & \textbf{Success Rate} & \textbf{Total Time} & \textbf{Avg} & \textbf{Total} & \textbf{Base} & \textbf{Additional} \\
\midrule
\textbf{GPT-5-mini} & 93.3\% $\pm$ 3.0\% & 23:21:02 & 0.03 & 7.44 & 76.5\% $\pm$ 5.1\% & 74.0\% $\pm$ 5.3\% \\
\textbf{GPT-5-nano} & 70.5\% $\pm$ 5.5\% & 40:40:01 & 0.02 & 6.05 & 65.0\% $\pm$ 5.7\% & 63.0\% $\pm$ 5.8\% \\
\textbf{GPT-OSS-120B} & 89.1\% $\pm$ 3.7\% & 21:32:12 & 0.02 & 6.03 & 81.5\% $\pm$ 4.7\% & 79.5\% $\pm$ 4.8\% \\
\textbf{Claude Haiku 4.5} & 94.0\% $\pm$ 2.9\% & 27:36:02 & 0.15 & 40.08 & 88.5\% $\pm$ 3.8\% & 86.0\% $\pm$ 4.2\% \\
\bottomrule
\end{tabular}%
\end{table}

\subsection{Large-Scale Evaluation and Maintenance}

\begin{table}[htpb]
\centering
\small
\caption{Large scale evaluation of NixBench with refinement disabled.}
\label{tab:full_dataset_results}
\begin{tabular}{l c c r r c c}
\toprule
& & & \multicolumn{2}{c}{\textbf{Cost (\$)}} & \multicolumn{2}{c}{\textbf{Automated Validation}} \\
\cmidrule(lr){4-5} \cmidrule(lr){6-7}
\textbf{Model} & \textbf{Success Rate} & \textbf{Total Time} & \textbf{Avg} & \textbf{Total} & \textbf{Base} & \textbf{Additional} \\
\midrule
\textbf{GPT-OSS-120B} & 89.3\% $\pm$ 3.7\% & 34:19:33 & 0.02 & 11.01 & 81.0\% $\pm$ 4.7\% & 79.0\% $\pm$ 4.9\% \\
\textbf{Claude Haiku 4.5} & 95.3\% $\pm$ 2.5\% & 44:46:42 & 0.14 & 66.47 & 88.5\% $\pm$ 3.8\% & 86.0\% $\pm$ 4.2\% \\
\bottomrule
\end{tabular}%
\end{table}

\begin{figure}[htbp]
     \centering
     \begin{subfigure}[b]{0.48\textwidth}
         \centering
         \includegraphics[width=\textwidth]{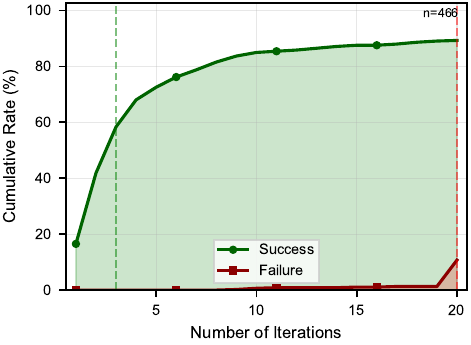}
         \caption{GPT-OSS-120b results.}
         \label{fig:rates_gpt}
     \end{subfigure}
     \hfill
     \begin{subfigure}[b]{0.48\textwidth}
         \centering
         \includegraphics[width=\textwidth]{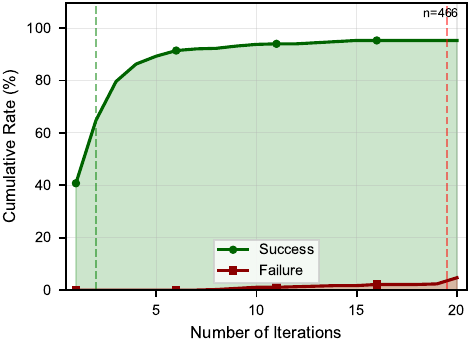}
         \caption{Claude Haiku 4.5 results.}
         \label{fig:rates_claude}
     \end{subfigure}
     \caption{Success and failure rates by iteration across 466 packaging tasks.}
     \label{fig:success_failure_rates_comparison}
\end{figure}

Next, we scaled our evaluation to the entire \textit{NixBench} dataset. However, of the 472 packaging tasks, Vibenix immediately failed to package 6 entries, resulting from issues on fetching and uncompressing the project source code to the builder machine during initial execution. Since these incompatibilities are scarce and do not represent proper execution of our solution, we excluded them from the large-scale evaluation.
Scaling to 466 packaging tasks, Claude Haiku 4.5 achieved 95.3\% build success at \$0.14 per package, while GPT-OSS-120B achieved 89.3\% at just \$0.02 per package. As shown in Figure~\ref{fig:success_failure_rates_comparison}, almost all successful builds resolve within the first five iterations. This plateau confirms the efficacy of the iterative loop while highlighting the impact that further research into early-stopping heuristics, such as progress evaluation, could have on mitigating the high cost of unproductive packaging efforts.

\subsection{Refinement and Runtime Correctness}

\begin{table}[htpb]
\centering
\small
\caption{Refinement-enabled run results.}
\label{tab:refinement_enabled_results}
\begin{tabular}{l c c r r c c}
\toprule
& & & \multicolumn{2}{c}{\textbf{Cost (\$)}} & \multicolumn{2}{c}{\textbf{Automated Validation}} \\
\cmidrule(lr){4-5} \cmidrule(lr){6-7}
\textbf{Model} & \textbf{Success Rate} & \textbf{Total Time} & \textbf{Avg} & \textbf{Total} & \textbf{Base} & \textbf{Additional} \\
\midrule
\textbf{GPT-OSS-120B} & 87.6\% $\pm$ 4.0\% & 46:16:09 & 0.06 & 16.42 & 83.5\% $\pm$ 4.5\% & 81.5\% $\pm$ 4.7\% \\
\textbf{Claude Haiku 4.5} & 92.9\% $\pm$ 3.1\% & 78:44:20 & 0.33 & 88.15 & 87.5\% $\pm$ 4.0\% & 85.0\% $\pm$ 4.3\% \\
\bottomrule
\end{tabular}%
\end{table}

Enabling the \textit{Refinement} step more than doubled execution time and cost. From our development testing, we expected a significant increase in resource costs, but this increase still went beyond our expectations.
Even more surprisingly, enabling this step had practically no impact on the automated validation metrics, which track functional correctness, the one aspect \textit{Refinement} should primarily improve.
The feature also had no impact on build success rates, but this was expected as it is only performed once a successful packaging is achieved. Table~\ref{tab:refinement_enabled_results} presents the results obtained.

One possible explanation for the automated validation results is that  some human-crafted validation checks are being ignored, or simply not recognized, during the build when the builder function of the generated package is distinct from its reference, such as incorrectly using \texttt{buildPythonApplication} for Python libraries. This allowed packages to build and skip certain validation checks, inflacting the pre-refinement results.
Consequently, rather than accepting these automated results \textit{de facto}, we viewed them as a hypothesis requiring further investigation. To simultaneously cross-examine these automated metrics and definitively assess the impact of the \textit{Refinement} step, we proceeded with manual validation of a 100-package subset of the generated results. Table~\ref{tab:refinement-validation} lists our results. 

\begin{table}[htbp]
\centering
\small
\caption{Manual VM validation for functional runtime correctness (n=100).}
\label{tab:refinement-validation}
\begin{tabular}{lrr}
\toprule
\textbf{Model} & \textbf{Raw (w/o Refinement)} & \textbf{Refined} \\
\midrule
\textbf{GPT-OSS-120B} & 51\% & 76\% \\
\textbf{Claude Haiku 4.5} & 47\% & 82\% \\
\bottomrule
\end{tabular}
\end{table}

Through this manual validation, we verified that, contrary to what the automated metrics report, \textit{Refinement} improved functional correctness by 25\% to 35\%. 
This outcome highlights the substantial impact of the \textit{Refinement} step, but also an existing limitation within automated validation, reaffirming the current necessity of manual verification to evaluate true functional correctness. 

\subsection{Maintenance}

We evaluate \textit{Vibenix}'s maintenance mode on a subset of \textit{NixBench-Repair}. Table~\ref{tab:maintenance_results} reports the results obtained. Both models achieved an impressive identical 87.6\% build success rate, although with high packaging time and costs, due to the \textit{Refinement} step being enabled. The average number of iterations to achieve a successful build remained consistent with the large-scale packaging validation (3.6 to 4.1), which alongside the high build success rate, highlights an identical challenge inherent to the two distinct tasks.

\begin{table}[htpb]
\centering
\small
\caption{Maintenance run results (with 2 step refinement).}
\label{tab:maintenance_results}
\begin{tabular}{l c c r r c c}
\toprule
& & & \multicolumn{2}{c}{\textbf{Cost (\$)}} & \multicolumn{2}{c}{\textbf{Automated Validation}} \\
\cmidrule(lr){4-5} \cmidrule(lr){6-7}
\textbf{Configuration} & \textbf{Success Rate} & \textbf{Total Time} & \textbf{Avg} & \textbf{Total} & \textbf{Base} & \textbf{Additional} \\
\midrule
\textbf{GPT-OSS-120B} & 87.6\% $\pm$ 4.0\% & 42:24:15 & 0.05 & 12.74 & 79.0\% $\pm$ 4.9\% & 76.3\% $\pm$ 5.1\% \\
\textbf{Claude Haiku 4.5} & 87.6\% $\pm$ 4.0\% & 50:47:21 & 0.25 & 66.57 & 81.6\% $\pm$ 4.7\% & 79.0\% $\pm$ 4.9\% \\
\bottomrule
\end{tabular}%
\end{table}

\subsection{Nix-ai-help}

Lastly, we evaluated the \texttt{nix-ai-help} tool on the \textit{NixBench} dataset. As previously mentioned, this packaging assistant lacks many of the features present on Vibenix. The evaluation of this assistant not only serves as a comparison of our approach against other existing solutions, but particularly as a comparison of our iterative build-repair loop against a baseline single-shot generation approach. Table~\ref{tab:nix_ai_help_results} presents the results obtained, which should be compared with those reported for Vibenix in Table~\ref{tab:full_dataset_results}.

\begin{table}[htpb]
\centering
\small
\caption{Results obtained using nix-ai-help.}
\label{tab:nix_ai_help_results}
\begin{tabular}{l c c r r c c}
\toprule
& & & \multicolumn{2}{c}{\textbf{Cost (\$)}} & \multicolumn{2}{c}{\textbf{Automated Validation}} \\
\cmidrule(lr){4-5} \cmidrule(lr){6-7}
\textbf{Configuration} & \textbf{Success Rate} & \textbf{Total Time} & \textbf{Avg} & \textbf{Total} & \textbf{Base} & \textbf{Additional} \\
\midrule
\textbf{GPT-OSS-120B} & 5.2\% $\pm$ 2.7\% & 01:17:59 & 0.0007 & 0.30 & 1.9\% $\pm$ 1.6\% & 1.9\% $\pm$ 1.6\% \\
\textbf{Claude Haiku 4.5} & 5.6\% $\pm$ 2.8\% & 00:58:28 & 0.0027 & 1.15 & 2.1\% $\pm$ 1.7\% & 1.9\% $\pm$ 1.6\% \\
\bottomrule
\end{tabular}%
\end{table}

Despite the dataset including 472 packaging tasks, \texttt{nix-ai-help} was only able to run on 427 entries. A portion of these stem from the reliance on the project web repository to obtain project information, and an incompatibility with certain remote repositories. Furthermore, packaging tasks in the dataset do not include a direct reference to their respective \textit{URL}, and so a dynamic resolution must be performed, which sometimes fails. We do not consider the missing 45 packages in the results presented in Table~\ref{tab:nix_ai_help_results}. 
Furthermore, a limited number of patches had to be done to \texttt{nix-ai-help}, primarily to allow support for the \textit{Bedrock} model provider and to output packaging costs. However, these changes do not impact the design or performance of the tool.  

Both models tested obtained low success rates, reaching about 5\% build success rate, compared to Vibenix's 80-90\% build success rate, demonstrating a substantial performance gap. 
From empirical observation, it appears that the packages generated often include a few simple lines which completely break evaluation and builds, namely due to package name resolution mistakes and missing key attributes.
On the upside, since the program does not provide the model with any tools, and only performs as single prompt, the packaging time and costs are greatly reduced, with both models averaging around an hour of total packaging time, and far below 0.01\$ per packaging task.

\section{Discussion}
\label{sec:discussion}
In this section, we answer our research questions based on the experimental results, discuss the limitations of our approach, and outline directions for future work.

\subsection{Answers to the Research Questions}
\paragraph{\textbf{RQ1: \RQOne}}
The high build success rates, as high as 95.3\% with Claude Haiku 4.5, validate that our iterative loop effectively grounds the LLM, allowing the agent to react to real-time compiler feedback.
Comparatively, our evaluation of single-shot generation with \texttt{nix-ai-help} resulted in only about 5\% packages successfully building, highlighting the importance and impact of the build-repair feedback loop.
However, achieving a successful compilation does not guarantee functional correctness. Our manual validation revealed that only roughly 47\% of raw building packages were functionally correct at runtime. This discrepancy confirms that while the agentic iterative loop excels at achieving successful builds, it requires additional context from runtime validation to ensure proper packaging. 

\paragraph{\textbf{RQ2: \RQTwo}}
The \textit{Refinement} step significantly contributes to bridging the compilation to runtime gap, improving functional correctness by up to 35\%. By allowing the agent to execute the package in a VM, it detects logical errors (such as packaging a Python library as an executable) that are invisible to static analysis. However, this comes at a steep penalty, increasing execution time and API expenses by as much as three times. Furthermore, the failure of our automated validation metric to capture these functional gains highlights inherent limitations with the metric, emphasizing the continued relevance of manual validation in determining true functional correctness.

\paragraph{\textbf{RQ3: \RQThree}}
The results from the 267 package maintenance subset confirm the extensibility of the iterative repair logic, as Vibenix succesfully built 87.6\% of projects with both GPT-OSS-120B and Claude Haiku 4.5. Vibenix demonstrated that autonomous agents can significantly reduce the manual burden of repository upkeep. However, results also stress the need for future development to lower \textit{Refinement}'s resource usage.

\subsection{Limitations}
\label{subsec:limitations}
While Vibenix represents a significant step toward automated hermetic packaging, it is designed as an AI agent to augment human effort rather than a total replacement for expert Nix maintainers. The scope of this work is bounded by several technical and ecosystem-related constraints that define the limits of autonomous intervention.

We categorize the primary limitations of the current system as follows:

\begin{enumerate}
    \item \textbf{Ecosystem Dependencies:} The success of the tool is inherently tied to the current state of \textit{Nixpkgs}. Vibenix will not autonomously package projects that rely on build tools or dependencies not yet available at the required version within the Nix ecosystem.
    
    \item \textbf{Build Impurity and Network Constraints:} Nix enforces a strict sandbox that prohibits network access during the build phase to ensure reproducibility. Projects that lack proper lock files or whose build scripts attempt to fetch assets from the internet will likely fail. While Vibenix attempts to resolve these via standard Nix fetchers, projects with \textit{impure} build processes remain outside the tool's autonomous reach.
    
    \item \textbf{Structural and Source Code Modifications:} In some instances, a project's source code is fundamentally incompatible with a hermetic environment and requires manual patching. While the agentic loop can identify these failures, performing source-level refactoring or patching to satisfy Nix's purity requirements is currently beyond the scope of this research.
    
    \item \textbf{Hardware and Platform:} Our evaluation focuses exclusively on standard Linux environments.
\end{enumerate}

By acknowledging these boundaries, we focus our research on projects that can be packaged within standard Nixpkgs workflows, which by themselves represent a significant manual burden for the Nix community.

\subsection{Future Work}
As the application of autonomous agents to hermetic package management is a relatively novel approach, there is much future work that can and should still be conducted on the topic.

Firstly, further iterations of Vibenix should explore optimizing its internal mechanisms, learning from the resource consumption and performance data obtained in this project. For example, it would be beneficial to conduct an in-depth analysis of the editing tools and progress evaluation modules to make them more efficient and generally applicable. 

Secondly, as has already been discussed, this work identified a significant discrepancy between automated build metrics and manual verification. Therefore, exploring and developing a more reliable metric for functional correctness would serve as a strong complement to this work, reducing the reliance on manual inspection.

Lastly, further evaluation of Vibenix and gathering real-world user feedback would encourage wider adoption and practical utility. For instance, conducting a large-scale evaluation specifically on maintenance tasks, and using a more complex dataset to assess the reliability of the dynamic release upgrade mechanism. Alongside this, a dedicated evaluation of the failure analysis step could be conducted to better understand its true accuracy in diagnosing root causes of failure. Additionally, integrating Vibenix into the workflows of active project developers and maintainers could provide a mutually beneficial feedback loop: practical usage would reveal real-world constraints and opportunities for improving Vibenix, while the packaging issues it exposes could motivate upstream fixes that make projects easier to package and consume more generally.

%% file: chapters/06_conclusion.tex
\section{Conclusion}
\label{sec:conclusion}

In this work, we introduced Vibenix, an AI agent for automated Nix package generation and maintenance, enhancing the scalability and accessibility of reproducible software distribution. By leveraging an extensive, domain-specific execution aware feedback loop, we have demonstrated that general purpose models can successfully navigate the challenges present in Nix packaging and maintenance. Our empirical evaluation using the newly developed NixBench and NixBench-Repair datasets shows that Vibenix produces buildable packages for as high as 95.3\% of cases, with manual validation confirming that about 80\% of these packages achieve functional correctness when the post build refinement step is employed. This research provides more than just a functional tool, as the curation of reproducible datasets and the validation of agentic repair logic offer a robust foundation for future work in automated software engineering. Ultimately, Vibenix serves to augment human expertise and reduce the manual burden on maintainers, bridging the gap between raw source code and verifiable distribution artifacts through a structured and automated approach.

%% file: chapters/07_acknowledgments.tex
\section*{Acknowledgments}
This work was supported by Fundação para a Ciência e a Tecnologia (FCT): N. Saavedra by grant BD/04736/2023;
M. Monis, N. Saavedra, and J.\,F. Ferreira by
project UID/50021/2025 (DOI: \href{https://doi.org/10.54499/UID/50021/2025}{\nolinkurl{https://doi.org/10.54499/UID/50021/2025}}), UID/PRR/50021/2025 (DOI: \href{https://doi.org/10.54499/UID/PRR/50021/2025}{\nolinkurl{https://doi.org/10.54499/UID/PRR/50021/2025}}); M. Monis, N. Saavedra, A. Mendes, and J.\,F. Ferreira by the following projects: 
the `InfraGov' project, with ref. n. 2024.07411.IACDC (DOI: 
\href{https://doi.org/10.54499/2024.07411.IACDC}{10.54499/2024.07411.IACDC}), funded by the `Plano de Recuperação e Resiliência (PRR)' under the investment `RE-C05-i08 - Ciência Mais Digital', measure `RE-C05-i08.M04' (in accordance with the FCT Notice No. 04/C05 i08/2024), framed within the financing agreement signed between the `Estrutura de Missão Recuperar Portugal (EMRP)' and the FCT as an intermediary beneficiary;
and 
the `SafeIac' project, co-funded by the European Union through the Lisboa 2030 Programme (ERDF) and by national funds through FCT, I.P., under project no. 15018 (DOI: \href{https://doi.org/10.54499/2023.18089.ICDT}{10.54499/2023.18089.ICDT}).

This work has also been supported by the JKU LIT Secure and Correct Systems Lab and Digidow, the Christian Doppler Laboratory for Private Digital Authentication in the Physical World.
We gratefully acknowledge financial support by the Austrian Federal Ministry of Economy, Energy and Tourism, the National Foundation for Research, Technology and Development, the Christian Doppler Research Association, 3 Banken IT GmbH, ekey biometric systems GmbH, Kepler Universit\"atsklinikum GmbH, NXP Semiconductors Austria GmbH \& Co KG, \"Osterreichische Staatsdruckerei GmbH, and the State of Upper Austria.

%% file: chapters/A_additional_validation.tex
\appendix

\section{Model Knowledge Cutoffs}
\label{sec:intended_application}

\begin{table}[h]
\centering
\small
\caption{Knowledge cutoff and release dates.}
\label{tab:model_cutoffs}
\begin{tabular}{lll}
\toprule
\textbf{Model} & \textbf{Knowledge Cutoff} & \textbf{Release Date} \\
\midrule
Claude Haiku 4.5 & Feb 2025\textsuperscript{a} & Oct 1, 2025\textsuperscript{b} \\
GPT-5-mini & May 31, 2024\textsuperscript{c} & Aug 7, 2025\textsuperscript{g} \\
GPT-5-nano & May 31, 2024\textsuperscript{f} & Aug 7, 2025 \\
GPT-OSS-120B & Jun 1, 2024\textsuperscript{h} & Aug 5, 2025\textsuperscript{j} \\
\bottomrule
\end{tabular}
\vspace{0.5em}
\small

\textsuperscript{a} \url{https://docs.anthropic.com/en/docs/about-claude/models/overview}

\textsuperscript{b} \url{https://www.anthropic.com/news/claude-haiku-4-5}

\textsuperscript{c} \url{https://platform.openai.com/docs/models/gpt-5-mini}

\textsuperscript{f} \url{https://platform.openai.com/docs/models/gpt-5-nano}

\textsuperscript{g} \url{https://openai.com/index/introducing-gpt-5/}

\textsuperscript{h} \url{https://platform.openai.com/docs/models/gpt-oss-120b}

\textsuperscript{j} \url{https://openai.com/index/introducing-gpt-oss/}
\end{table}

\subsection{Data Contamination Prevention}

All knowledge cutoff dates presented in Table~\ref{tab:model_cutoffs} precede the commits in NixBench, preventing training data contamination.
Anthropic is the only provider publishing a post-training and fine-tuning cutoff date (July 2025), providing additional assurance against contamination.
We deemed the one-week overlap between OpenAI model releases (August 5-7, 2025) and our dataset start (August 1, 2025) acceptable, assuming fine-tuning datasets would not change in a relevant manner one week before release.